\documentclass[12pt,preprint]{aastex}
\usepackage{pslatex}
\begin{document}

\title{Thermal conduction and particle transport in strong MHD
turbulence, with application to galaxy-cluster plasmas}
\author{Benjamin D. G. Chandran } \affil{Department of Physics \&
Astronomy, University of Iowa} \email{ benjamin-chandran@uiowa.edu}
\author{Jason L. Maron} \email{jason-maron@uiowa.edu}
\affil{Department of Physics \& Astronomy, University of Iowa, and
Department of Physics \& Astronomy, University of Rochester}

\begin{abstract}
Galaxy clusters possess turbulent magnetic fields with a dominant
scale length~$l_B \simeq 1-10$~kpc.  In the static-magnetic-field
approximation, the thermal conductivity~$\kappa_T$ for heat transport
over distances~$\gg l_B$ in clusters is~$\simeq \kappa_{\rm S}
l_B/L_{\rm S}(\rho_e)$, where $\kappa_{\rm S}$ is the Spitzer thermal
conductivity for a non-magnetized plasma, the length~$L_{\rm S}(r_0)$
is a characteristic distance that a pair of field lines separated by a
distance~$r_0<l_B$ at one location must be followed before they
separate by a distance~$l_B$, and~$\rho_e$ is the electron gyroradius.
We introduce an analytic Fokker-Planck model and a numerical Monte
Carlo model of field-line separation in strong magnetohydrodynamic
(MHD) turbulence to calculate~$L_{\rm S}(r_0)$. We also
determine~$L_{\rm S}(r_0)$ using direct numerical simulations of MHD
turbulence with zero mean magnetic field.
All three approaches, like earlier models, predict that
$L_{\rm S}$ asymptotes to a value of order several~$l_B$ as $r_0$ is
decreased towards~$l_d$ in the large-$l_B/l_d$ limit, where $l_d$ is
the dissipation scale, which is taken to be the proton
gyroradius. When the turbulence parameters used in the Fokker-Planck
and Monte Carlo models are evaluated using direct numerical
simulations, the Fokker-Planck model yields $L_{\rm S} (\rho_e) \simeq
4.5 l_B$ and the Monte Carlo model yields $L_{\rm S}(\rho_e) \simeq
6.5 l_B$ in the large-$l_B/l_d$ limit.  Extrapolating from our direct
numerical simulations to the large-$l_B/l_d$ limit, we find
that~$L_{\rm S}(\rho_e) \simeq  5-10 l_B$, implying that $\kappa_T
\simeq 0.1\kappa_{\rm S} - 0.2\kappa_{\rm S}$ in galaxy clusters in the
static-field approximation.  We also discuss the phenomenology of
thermal conduction and particle diffusion in the presence of
time-varying turbulent magnetic fields. Under the questionable
assumption that turbulent resistivity completely reconnects field
lines on the time scale~$l_B/u$, where $u$ is the rms turbulent
velocity, we find that~$\kappa_T$ is enhanced by a moderate amount
relative to the static-field estimate for typical cluster conditions.
\end{abstract}
\maketitle

\section{Introduction}
\label{sec:intro} 

In the cooling-flow (CF) model of intracluster plasma, radiative
cooling causes plasma to flow towards a cluster's
center and cool to sub-x-ray temperatures, presumably ending up as
either stars, smaller compact objects, and/or cold gas. Aside from the
gravitational work done on inflowing plasma, heating of intracluster
plasma is neglected in the model, and mass accretion rates $\dot{M}$
are as high as $10^3 M_\sun \mbox{ yr}^{-1}$ for some clusters (Fabian
1994). A longstanding problem for the CF model has been the difficulty
in accounting for all the accreted mass.  For example, the observed
rates of massive star formation are a factor of 10-100 less than
expected if the cooling plasma predicted by the model ends up forming
stars with a normal IMF (Crawford et~al 1999, Fabian 2002).  In
addition, recent x-ray observations find no evidence of plasma cooling
to temperatures below 1-2~keV (Peterson et~al 2001, Tamura et~al
2001). These difficulties suggest that some form of heating
approximately balances radiative cooling, thereby dramatically
reducing~$\dot{M}$ relative to CF estimates.  A number of heating
mechanisms have been considered, such as galaxy motions (Bregman \&
David 1989), supernovae (Bregman \& David 1989), cosmic-rays
(Bohringer \& Morfill 1988, Tucker \& Rosner 1983), active galactic
nuclei (Pedlar et~al 1990, Tabor \& Binney 1993, Binney \& Tabor 1995,
Ciotti \& Ostriker 2001, Churazov et~al 2002), 
dissipation of turbulent energy (Loewenstein \& Fabian 1990,
Churazov et~al 2003, Chandran 2003), and thermal
conduction, which can transport heat from the hot outer regions of a
cluster into the relatively cooler core (Binney \& Cowie 1981,
Tribble 1989, Tao 1995, Chandran \& Cowley 1998,
Narayan \& Medvedev 2001, Gruzinov 2002,
Voigt et~al 2002, Zakamska \& Narayan 2002).  For thermal conduction
to approximately balance cooling, the thermal conductivity~$\kappa_T$
must be a significant fraction of the Spitzer value for a
non-magnetized plasma,~$\kappa_{\rm S}$, and in some clusters even
greater than~$\kappa_{\rm S}$~(Fabian 2002, Zakamska \& Narayan 2002),
where
\begin{equation}
\kappa_{\rm S} = 5.2 \times 10^{32} \left(\frac{k_B T}{\mbox{10 keV}}\right)^{5/2}
\left( \frac{10^{-3}\mbox{ cm}^{-3}}{n_e}\right)\left(\frac{37}{\ln \Lambda}
\right)\; \frac{\mbox{cm}^2}{\mbox{s}},
\label{eq:kappaS} 
\end{equation} 
$T$ is the temperature, $k_B$ is the Boltzmann constant,
$n_e$ is the electron density, and~$\ln \Lambda$ is the Coulomb
logarithm (Spitzer 1962).\footnote{In terms of this definition,
in which~$\kappa_{\rm S}$ is expressed as a diffusion coefficient,
the heat flux is given by~$-n_e k_B \kappa_{\rm S}\nabla T$.}

Galaxy clusters are filled with tangled magnetic fields with a
dominant length scale~$l_B \simeq 1-10$~kpc (Kronberg 1994, Taylor et~al
2001, 2002) that is much less than the size of a cluster
core,~$R_c\simeq 100$~kpc. Both optical-line-emitting gas in clusters
and hot intracluster plasma are observed to be in turbulent motion
(Fabian 1994, Churazov et~al 2003).  The effects of turbulent magnetic
fields and velocities on~$\kappa_T$ are the subject of this paper, and
have been investigated by a number of authors (e.g., Tribble 1989, Tao
1995, Chandran \& Cowley 1998, Chandran et~al 1999, Narayan \&
Medvedev 2001, Malyshkin \& Kulsrud 2001, Gruzinov 2002).

Transport in the presence of strong turbulence is a difficult and
unsolved problem. It is likely that
the thermal conductivity for a particle species scales like the
test-particle diffusion coefficient for that particle species
(Rechester \& Rosenbluth 1978, Krommes, Oberman, \& Kleva 1983):
since the collisional transfer of energy between particles occurs locally in
space,  diffusion of heat accompanies the diffusion of
heat-carrying particles. We thus
estimate~$\kappa_T$ in clusters from the relation
\begin{equation}
\frac{\kappa_T}{\kappa_{\rm S}} \simeq \frac{D}{D_0},
\label{eq:defkt} 
\end{equation} 
where $D$ is the diffusion coefficient of thermal electrons
 and $D_0$ is the thermal-electron
diffusion coefficient in a non-magnetized plasma.
\footnote{Electrons generally make the dominant contribution
to~$\kappa_T$ because they diffuse more rapidly than ions.}
\footnote{In a steady state, an electric field
is set up to maintain quasi-neutrality.
This electric field reduces $\kappa_T$ by a factor of~$\simeq 0.4$ in a
non-magnetized plasma (Spitzer 1962). We do not consider the effects
of turbulence on this reduction factor.} 
\footnote{There is
some ambiguity in the right-hand side of equation~(\ref{eq:defkt}). We
take $D$ and $D_0$ to be diffusion 
coefficients for particles of a specified energy as opposed to diffusion
coefficients of a particle that diffuses in energy as well as space. Thus,
$D/D_0$ may scale differently for electrons of
energies, e.g., $k_B T$ and~$2k_B T$.
In the static-field
approximation this is not an issue since $D/D_0$ is the same for all
energies of interest for cluster parameters.  However, when we
estimate the effects of turbulent resistivity, we pick a
representative energy at which to evaluate $D/D_0$ as discussed in
section~\ref{sec:rec}.}

Since thermal electrons in clusters move much faster than the ${\bf E}
\times {\bf B} $ velocity of field lines, a reasonable first
approximation for~$D$ is obtained by treating the magnetic field as
static.  In a collisional plasma, particle diffusion over
distances~$\gg l_B$ in a static field depends critically on the rate
of separation of neighboring magnetic field lines (Rechester \&
Rosenbluth 1978, Chandran \& Cowley 1998).  If two field lines within
a snapshot of strong magnetohydrodynamic (MHD) turbulence are at some
location separated by a distance~$r_0 \ll l_B$, they will separate by
a distance~$l_B$ after some distance~$z$ along the magnetic field. The
length~$L_{\rm S}(r_0)$ is a characteristic
value of~$z$ defined in section~\ref{sec:static}.  In the
static-magnetic field approximation, the thermal
conductivity~$\kappa_T$ in galaxy clusters over distances $\gg l_B$
satisfies
\begin{equation}
\kappa_T \simeq\frac{\kappa_{\rm S} l_B}{L_{\rm S}(\rho_e)},
\label{eq:kt0} 
\end{equation} 
where $\rho_e$ is the electron gyroradius. Equation~(\ref{eq:kt0})
makes use of the fact that electron motion along the magnetic
field is relatively unimpeded by magnetic mirrors and whistler
waves excited by the heat flux because the Coulomb mean free path
is short compared to both~$l_B$ and the temperature-gradient length
scale (see section~\ref{sec:dpar}).
The factor $l_B/L_{\rm S}(\rho_e)$ in
equation~(\ref{eq:kt0}) measures the reduction in~$\kappa_T$ 
associated with tangled field lines, which increase the distance
electrons must travel in going from hotter regions to colder regions.

Three previous studies (Jokipii~1973, Skilling, McIvor, \&
Holmes~1974, Narayan \& Medvedev~2001) calculated~$L_{\rm S}(r_0)$ for
strong turbulence assuming a power spectrum of magnetic fluctuations
on scales ranging from~$l_B$ to a much smaller dissipation
scale~$l_d$, which is approximately the proton gyroradius~$\rho_i$ in
galaxy clusters (Quataert 1998). Using different definitions
of~$L_{\rm S}$, each study found that~$L_{\rm S}(r_0)$ is of order~$l_B$ in
the small-$(r_0/l_B)$ and small-$l_d/l_B$ limits provided $r_0$ is not
vastly smaller than~$l_d$. Jokipii~(1973) assumed an isotropic
turbulent magnetic field and employed a stochastic model of field-line
separation, which he solved using a Monte Carlo numerical method.
Skilling et~al~(1974) assumed isotropic turbulence and introduced an
approximate equation of the form~$d\langle r\rangle /dz = F(\langle
r\rangle)$ for the average separation~$r$ of a pair of field lines,
with the function~$F$ estimated from the power spectrum of the
turbulence.  Narayan \& Medvedev~(2001) introduced a similar equation
for the evolution of the mean square separation, $d\langle r^2
\rangle/dl = G(\langle r^2\rangle)$, and estimated $G$ using the
Goldreich-Sridhar~(1995)~model of anisotropic MHD turbulence.

In this paper, we calculate~$L_{\rm S}$ using three different
methods.  First, we consider an approximate model 
in which the separation~$r$ of a pair of field lines evolves
stochastically and is described by the Fokker-Planck equation
\begin{equation}
\frac{\partial P}{\partial l} = - \frac{\partial }{\partial y}
\left[\left(\frac{\langle \triangle y \rangle}{\triangle l}\right)
\,P\right] + \frac{\partial^2}{\partial y^2}\left[
\left(\frac{ \langle (\triangle y)^2 \rangle}{2\triangle l}\right) \, P\right],
\label{eq:fp0a} 
\end{equation} 
where $y = \ln (r/l_B)$, $l$ is distance along the field in units
of~$l_B$, $\triangle y$ is the increment to $y$ after a field-line
pair is followed a distance $\triangle l$ along the field, $\langle
\dots \rangle$ is an average over a large number of field-line pairs,
and $P(y_1,l)dy$ is the probability that $y$ is in the interval
$(y_1,y_1+dy)$ after a distance~$l$ along the field. Our model is
similar to Jokipii's (1973), although we determine the functional form
of $\langle \triangle y \rangle /\triangle l$ and $\langle (\triangle
y)^2 /\triangle l$ using the Goldreich-Sridhar model of locally
anisotropic MHD turbulence,
and we solve the Fokker Planck equation analytically, which allows us
to determine the functional dependence of~$L_{\rm S}$ on turbulence
parameters. If we evaluate the turbulence parameters in the model
using data from direct numerical simulations, we find that~$L_{\rm
S}(r_0)\rightarrow 4.5 l_B$ as~$r_0$ is
decreased towards~$l_d$ in the large-$l_B/l_d$ limit,
and that $L_{\rm S}(\rho_e) \simeq L_{\rm S}(l_d)$.

Our second method for calculating $L_{\rm S}(r_0)$ uses a numerical
Monte Carlo model of field line separation in which each
random step $\triangle y$ is of order unity (the Fokker-Planck equation
assumes infinitesimal Markovian steps). When the model parameters
are evaluated using data from direct numerical simulations,
the Monte Carlo model gives~$L_{\rm S} (\rho_e) 
\simeq 6.5 l_B$ in the large-$l_B/l_d$ limit.

Our third method for calculating $L_{\rm S}(r_0)$ involves tracking
field-line trajectories in direct numerical simulations of MHD
turbulence with zero mean magnetic field. The results of our highest
resolution simulations are consistent with the prediction of the
theoretical studies that $L_{\rm S}(r_0)$ approaches a value of order
several~$l_B$ as~$r_0$ is decreased towards~$l_d$ in the
large-$l_B/l_d$ limit. Extrapolating our numerical results to the
large-$l_B/l_d$ limit suggests that~$L_{\rm S}(\rho_e) \simeq 5-10
l_B$ in clusters.

Field evolution and turbulent fluid motions increase electron (and
ion) mobility, enhancing~$\kappa_T$ to some degree.  Turbulent
diffusion in clusters has been studied by Cho et~al (2003).  In this
paper we develop a phenomenology to describe the interplay between
field evolution and single-electron motion under the questionable 
assumption that turbulent resistivity completely reconnects
field lines on the time scale~$l_B/u$, where $u$ is the rms turbulent
velocity. A similar assumption was explored by Gruzinov (2002).  We
find three limiting cases for the thermal conductivity.  For 
typical cluster parameters and turbulent velocities (Churazov et~al 2003)
 $\kappa_T\sim \sqrt{\kappa_{\rm S} u
l_B}$, a value that is somewhat larger than both the turbulent diffusivity~$\sim u
l_B$ for that cluster and the static-field estimate of $\kappa_{\rm S}
l_B/L_{\rm S} \simeq 0.1\kappa_{\rm S}-0.2\kappa_{\rm S}$.  More work,
however, is needed to clarify the role of turbulent resistivity before
firm conclusions can be drawn about its effects on~$\kappa_T$.
Additional work is also needed to quantify factors of order unity that
have been neglected in estimating~$\kappa_T$ both in the presence of
efficient turbulent resistivity and in the static-field approximation.

The remainder of this paper is organized as follows. In
section~\ref{sec:static} we review the phenomenology of thermal
conduction in static tangled magnetic fields. In
section~\ref{sec:dpar}, we discuss the effects of magnetic mirrors and
microturbulence on electron diffusion along field lines. We present
our Fokker-Planck and Monte Carlo models of field-line separation in
section~\ref{sec:FP}. We compare these approximate theoretical models
with results from direct numerical simulations in
section~\ref{sec:comp}.  We estimate the effects of turbulent
resistivity on $\kappa_T$ in section~\ref{sec:rec} and summarize our conclusions in
section~\ref{sec:conc}. We present numerical simulations of field-line
separation for different types of MHD turbulence in a companion paper
(Maron, Chandran, \& Blackman 2003).  Table~\ref{tab:t1} defines some
frequently used notation.

\begin{table*}[h]
\begin{tabular}{lc}
\hline
\hline 
&
\vspace{-0.25cm} 
\\
Notation & Meaning \vspace{0.1cm} \\
\hline  
\vspace{-0.2cm} \\
$l_B$ & dominant length scale of the magnetic field \\
$l_d$ & magnetic dissipation scale  \\
$\langle l \rangle $  & average distance in units of $l_B$ that a field-line pair  must be followed before separating by a distance $l_B$ \\
$\langle l^2 \rangle $  &  average square of the distance
 in units of $l_B^2$ that a field-line pair must be followed before separating by a distance $l_B$ \\
$L_{\rm S}(r_0)$ & $l_B\langle l^2 \rangle /\langle l \rangle$---characteristic 
distance a pair of field lines separated by a distance~$r_0$ must be followed before separating to~$l_B$ \\
$\rho_e$ & electron gyroradius \\
$\kappa_T$ & thermal conductivity \\
$\kappa_{\rm S}$ & Spitzer thermal conductivity in a non-magnetized plasma \\
$D$ & 3D single-electron diffusion coefficient \\
$D_0$ & 3D single-electron diffusion coefficient in a non-magnetized plasma \\
$D_\parallel$ & diffusion coefficient for electron motion along the magnetic field \\
$\lambda$ & thermal-electron Coulomb mean free path \\
$u$ & rms turbulent velocity \vspace{0.2cm} \\
\hline
\hline
\end{tabular}
\caption{Definitions.
\label{tab:t1} }
\end{table*}

\section{The phenomenology of thermal conduction in a static tangled magnetic field}
\label{sec:static} 

We assume that the magnetic fluctuations possess an
inertial range extending from an outer scale $l_B$ to a much smaller
inner scale $l_d$ with the magnetic energy dominated by scales $\simeq
l_B$. Except where specified, the discussion focuses on the case
relevant for clusters in which the mean magnetic field
is negligible.

A tangled magnetic field line is essentially a
random-walk path through space.  If a particle is tied to a single
field line and travels a distance $l\gg l_B$ along the static magnetic
field, it takes $\sim l/l_B$ random-walk steps of length $\sim l_B$, resulting
in a mean-square three-dimensional displacement of
\begin{equation}
\langle (\triangle x)^2 \rangle = \alpha l_B l,
\label{eq:tr} 
\end{equation} 
where $\alpha$ is a constant of order unity (values of 
$\alpha$ for the numerical simulations used in this paper
are listed in table~\ref{tab:t2}).  When there is a mean
field ${\bf B} _0$ comparable to the rms field, $\langle (\triangle
x)^2 \rangle$ in equation~(\ref{eq:tr}) is interpreted as the
mean-square displacement perpendicular to ${\bf B}_0$. If the
particle's motion along the field is diffusive with diffusion
coefficient $D_\parallel$, then
\begin{equation}
l \sim \sqrt{D_\parallel t},
\label{eq:l1} 
\end{equation} 
and~(Rechester \& Rosenbluth 1978, Krommes et~al 1983)
\begin{equation}
\langle (\triangle x)^2 \rangle \propto t^{1/2},
\label{eq:t2} 
\end{equation} 
indicating subdiffusion: $D\equiv \displaystyle\lim_{t \rightarrow \infty}
\langle(\triangle x)^2 \rangle/6t \rightarrow 0$ (see also
Qin et~al 2002a,b).  This process is
called double diffusion: the particle diffuses along the field line,
and the field line itself is a random-walk path through space.

\begin{figure*}[h]
\vspace{9cm}
\includegraphics{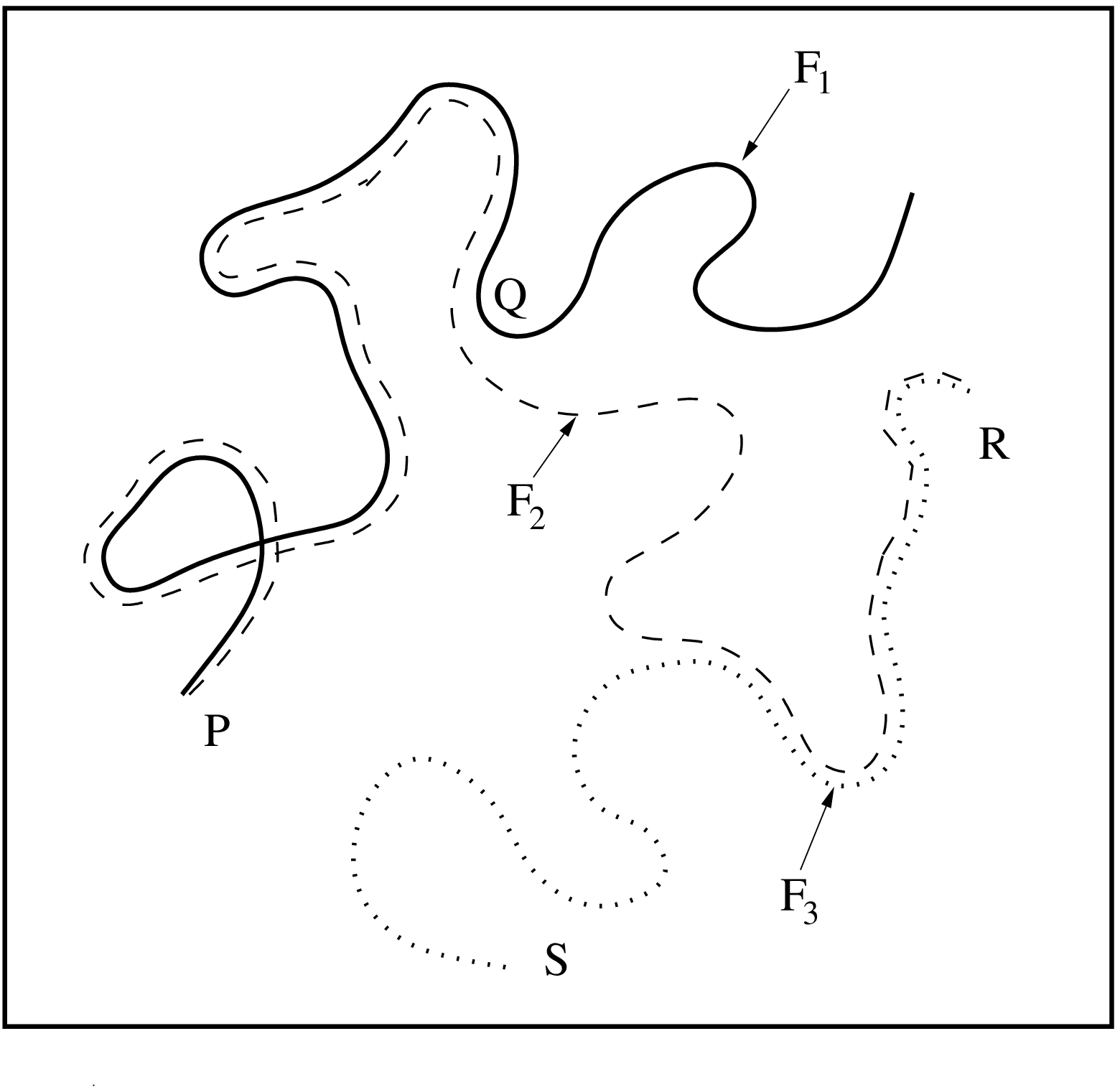}
\caption{Trajectory of a diffusing electron.
\label{fig:f1}}
\end{figure*}

The vanishing of~$D$ for a particle tied to a single field line can be
understood by considering a particle starting out at point~P in
figure~\ref{fig:f1}. If this particle moves one Coulomb mean free
path~$\lambda$ along its field line~F$_1$ towards point Q and then
randomizes its velocity due to collisions, it has a $\sim 50$\% chance of
changing its direction of motion along the magnetic field and
returning to its initial point~P. In contrast, in a Markovian three
dimensional random walk, the second step is uncorrelated from the
first and there is a vanishing probability that a particle will return
to its initial location.  In a cluster, $\rho_e/l_B \simeq 10^{-15}$,
and thus it is tempting to assume that electrons are tied to field
lines. The importance of equation~(\ref{eq:t2}) is that any study that
assumes that electrons are perfectly tied to field lines will conclude
that~$\kappa_T= D=0$.

Of course, an electron is not tied to a single field line. As pointed
out by Rechester \& Rosenbluth (1978), small cross-field motions
enhanced by the divergence of neighboring field lines leads to
a non-zero~$D$. This can be seen with the aid of
figure~\ref{fig:f1}.
Suppose an electron starts out at point~P on field line~F$_1$
traveling towards point~Q. After moving a short distance, field
gradients and collisions cause the particle to take a step of
length~$\sim\rho_e$ across the magnetic field, from~F$_1$ to a new
field line~F$_2$. Although the electron continuously drifts across the
field to new field lines, let us assume for the moment that it remains
attached to~F$_2$.  As the electron follows~F$_2$, F$_2$ diverges from
F$_1$. Let $z$ be the distance that F$_2$ must be followed before
F$_2$ separates from F$_1$ by a distance~$l_B$. (Because the electron
continuously drifts across the field, it typically separates from F$_1$
after traveling a distance somewhat less than~$z$ along the field;
this effect, however, is ignored in this paper.)  After the electron
moves a distance~$z$ along F$_2$, its subsequent motion is not
correlated with~$F_1$. The electron proceeds to point~R, and then its
collisional random walk along the magnetic field changes direction,
bringing it back towards point~Q. Instead of following F$_2$ back to
point~Q, however, the electron drifts across the field and ends up on
a new field line~F$_3$. After following F$_3$ for a distance~$\sim z$,
the electron separates from F$_2$ by a distance~$\sim l_B$ and
proceeds to point~S.

In this example, the electron's small cross-field motions and
the divergence of neighboring field lines allow the electron to
escape from its initial field line and undergo a Markovian random walk
in three dimensions. The fundamental random walk step is a
displacement of length $mz$ along the magnetic field, where~$m$ is
some constant of order unity, perhaps 2 or~3.  From equation~(\ref{eq:tr}), when $mz
\gg l_B$, a single random step corresponds to a 3D displacement of
\begin{equation}
(\triangle x)^2 \sim \alpha mz l_B.
\label{eq:tr3} 
\end{equation} 
[When $mz \gg l_B$, the difference between the actual value
of $(\triangle x)^2$ and its mean becomes small.]
When $mz \gg \lambda$, where $\lambda$ is the Coulomb mean free path,
a single step takes a time
\begin{equation}
\triangle t  \sim \frac{ m^2 z^2}{D_\parallel}.
\label{eq:dt} 
\end{equation} 
When $mz$ is only moderately greater than $l_B$ or $\lambda$,
equations~(\ref{eq:tr3}) and (\ref{eq:dt}) remain approximately
valid. During successive random steps, a particle will find itself in
regions of differing magnetic shear, and thus $z$ will vary. The
diffusion coefficient is given by $D= \langle (\triangle x)^2
\rangle/6\langle \triangle t \rangle$ where $\langle \dots \rangle$ is
an average over a large number of steps~(Chandrasekhar 1943). Ignoring
factors of order unity we obtain
\footnote{In applying equation~(\ref{eq:tr3}) first and then averaging $z$
and~$z^2$, we are averaging separately over the wandering of
a single field line through space and the separation of neighboring
field lines. This is justified to some extent since the former
is dominated by the value of~${\bf B}$ at the outer scale and
the latter depends on the magnetic shear throughout the inertial
range. Although some error is introduced by averaging separately,
equations~(\ref{eq:d2}) and (\ref{eq:deflrr}) are sufficient
for estimating~$\kappa_T$.}
\begin{equation}
D \sim \frac{ D_\parallel l_B}{L_{\rm S}},
\label{eq:d2} 
\end{equation} 
with
\begin{equation}
L_{\rm S} = \frac{\langle z^2\rangle}{\langle z \rangle}.
\label{eq:deflrr} 
\end{equation} 
In general, $L_{\rm S}$, $\langle z \rangle$, and $\langle z^2 \rangle$
are functions of the initial separation of a field-line pair,~$r_0$,
but for electron thermal conduction we have set~$r_0 = \rho_e$.
If there is a mean magnetic field comparable to the fluctuating field,
equation~(\ref{eq:d2}) is recovered provided $D$ is replaced by
$D_\perp$, the coefficient of diffusion perpendicular to the mean
field. We note that taking $L_{\rm S} = \langle z \rangle $ leads to similar
results since $\langle z \rangle  \sim \langle z^2\rangle /\langle z \rangle$ in our
direct numerical simulations and Fokker-Planck and
Monte Carlo calculations. 
In clusters $D_\parallel \sim D_0$ as discussed in
section~\ref{sec:dpar}, where~$D_0$ is the electron diffusion
coefficient in a non-magnetized plasma. As a result,
equations~(\ref{eq:defkt}) and (\ref{eq:deflrr}) give
\begin{equation}
\kappa_T \simeq \frac{\kappa_S l_B}{L_{\rm S}}.
\label{eq:kts2} 
\end{equation}

\section{Electron diffusion along the magnetic field}
\label{sec:dpar} 

For mono-energetic electrons subject to a fixed Coulomb pitch-angle
scattering frequency, the diffusion coefficient~$D_{\parallel,0}$ for
motion along a uniform magnetic field is equal to the
three-dimensional diffusion coefficient~$D_0$ for motion in a
non-magnetized plasma.  Two mechanisms for suppressing $D_\parallel$
relative to~$D_{\parallel,0}$ have been discussed in the literature:
magnetic mirrors and wave pitch angle scattering. Magnetic mirrors
associated with a cluster's turbulent magnetic field significantly
reduce~$D_\parallel$ only when $\lambda\gtrsim l_B$, where $\lambda$
is the Coulomb mean free path of thermal electrons (Chandran \& Cowley
1998, Malyshkin \& Kulsrud 2001). In cluster cores, however, $\lambda$
is significantly less than~$l_B$, and thus mirrors have only a small
effect.  [For Hydra A and 3C295, $\lambda \sim 1 $~kpc at 100~kpc from
cluster center, and $\lambda = 0.02-0.05$~kpc at 10~kpc from cluster
center (Narayan \& Medvedev 2001).]  When the Knudsen number~$N_K
\equiv \lambda/L_{T,\parallel}$ approaches~1, where $L_{T,\parallel} =
T/|\hat{\bf b} \cdot \nabla T|$ and $\hat{b}$ is a unit vector
pointing along the magnetic field, the heat flux becomes large and
excites whistler waves that enhance the pitch-angle scattering of
electrons and reduce~$D_\parallel$ (Pistinner \& Eichler
1998)\footnote{The formula in Pistinner \& Eichler's (1998) paper is
$\kappa_T/\kappa_S \sim 1/[1 + 250\beta_e(\lambda/L_{T,\parallel})]$, but the
factor of 250 should be corrected to a factor of 10 (Pistinner,
private communication)}.  However, for heat conduction into a cluster
core, $|\nabla T| \sim T/R_c$ with $R_c \sim 100$~kpc, and $N_K \ll
1$.  Thus, wave pitch-angle scattering has only a small effect, and
\begin{equation}
D_\parallel\sim D_0. 
\label{eq:dpard0} 
\end{equation} 
We note that the chaotic trajectories of field lines cause
$L_{T,\parallel} $ to be larger than $T/|\nabla T|$.

\section{Fokker-Planck and Monte Carlo models of field-line separation in strong
MHD turbulence}
\label{sec:FP}

We adopt the Goldreich \& Sridhar (1995) 
model of strong locally anisotropic MHD
turbulence, which is supported by direct numerical
simulations (Cho \& Vishniac 2000, Maron \& Goldreich 2001,
Cho \& Lazarian 2003). We assume that the fluctuating field
is equal to or greater than any mean field in the system.  The separation of
neighboring magnetic field lines in strong MHD turbulence is dominated
by shear Alfv\'en modes.  On scales smaller than~$l_B$, an
Alfv\'en-mode eddy is elongated along the direction of the average of
the magnetic field within the volume of the eddy, denoted ${\bf B}
_{\rm local}$, with (Goldreich \& Sridhar 1995,
Cho \& Vishniac 2000, Maron \& Goldreich 2001,
Lithwick \& Goldreich 2001)
\begin{equation}
 B_{\lambda_\perp}\sim B_{\rm local} \left(\frac{\lambda_\perp}{l_B}\right)^{1/3},
\label{eq:u0} 
\end{equation} 
and
\begin{equation}
\lambda_\parallel \sim \lambda_\perp^{2/3} l_B^{1/3},
\label{eq:lperp} 
\end{equation} 
where $B_{\lambda_\perp}$ is the rms magnetic fluctuation of an
Alfv\'en-mode eddy of width $\lambda_\perp$ measured across ${\bf B}
_{\rm local}$ and length $\lambda_\parallel$ measured along ${\bf
B}_{\rm local}$. In fully ionized plasmas, the dissipation scale~$l_d$
for Alfv\'en modes is set by collisionless effects, and is comparable
to the proton gyroradius~$\rho_i$ (Quataert 1998).
The magnetic-field perturbation of an Alfv\'en mode is perpendicular
to~${\bf B} _{\rm local}$.  Equations~(\ref{eq:u0}) and
(\ref{eq:lperp}) thus imply that when two field lines separated by a
distance~$r$ are followed for a distance~$r^{2/3}l_B^{1/3}$,
$r$~either increases or decreases by a factor of order unity assuming
$l_d< r < l_B$~(Narayan \& Medvedev 2001, Maron \& Goldreich 2001, Lithwick
\& Goldreich 2001). If $r < l_d$, the separation or
convergence of the field lines is dominated by the eddies of
width~$l_d$, and $r$ increases or decreases by a factor of order unity
when the field lines are followed a distance~$l_d^{2/3} l_B^{1/3}$
(Narayan \& Medvedev 2001). We define
\begin{equation}
\triangle l =\left\{ \begin{array}{cc}
(r/l_B)^{2/3} & \mbox{ if $l_d< r < l_B$} \\
(l_d/l_B)^{2/3} & \mbox{ if $r<l_d$} 
\end{array} \right.,
\label{eq:defdell} 
\end{equation} 
\begin{equation}
y = \ln(r/l_B),
\label{eq:defy} 
\end{equation} 
$\triangle y$ to be the change in~$y$ when the field lines
are followed a distance~$\triangle l$,
\begin{equation}
a = \langle \triangle y \rangle,
\label{eq:defa} 
\end{equation}
with $a$ taken to be positive, and
\begin{equation}
b = \langle (\triangle y)^2 \rangle/2,
\label{eq:defb} 
\end{equation}
where $a$ and $b$ are of order unity.

To obtain Monte Carlo and analytic solutions for~$L_{\rm S}$, we make
several approximations.  The changes in~$r$ over a distance~$r^{2/3}
l_B^{1/3}$ along the field associated with eddies of width much
smaller or larger than~$r$ (or~$l_d$ if $r< l_d$) are small compared to
the changes arising from eddies of width~$r$ (or~$l_d$ if
$r<l_d$) and are neglected.  We also take $a$ and $b$ to be
independent of $l$ and $y$, and consecutive values of~$\triangle y$ to
be uncorrelated.  To obtain an approximate analytic solution
for~$L_{\rm S}$, we make the further approximation of describing the
stochastic variation of~$y$ with the Fokker-Planck equation
\begin{equation}
\frac{\partial P}{\partial l} = - \frac{\partial }{\partial y}
\left[\left(\frac{\langle \triangle y \rangle}{\triangle l}\right)
\,P\right] + \frac{\partial^2}{\partial y^2}\left[
\left(\frac{ \langle (\triangle y)^2 \rangle}{2\triangle l}\right) \, P\right],
\label{eq:fp0} 
\end{equation} 
where $P(y^\prime,l)dy$ is the probability that~$y$
 is in the interval
$(y^\prime, y^\prime+dy)$, and $l$ is distance
along the magnetic field in units of~$l_B$. The additional
approximation in introducing equation~(\ref{eq:fp0}) is associated
with~$y$ changing by order unity during a
single random step [equation~(\ref{eq:fp0}) assumes infinitesimal
steps].

We now solve equation~(\ref{eq:fp0}) to obtain an analytic solution
for~$L_{\rm S}$.  Substituting equations~(\ref{eq:defdell}), (\ref{eq:defa}),
and~(\ref{eq:defb}) into equation~(\ref{eq:fp0}) yields
\begin{equation}
\frac{\partial P}{\partial l} = - \frac{\partial \Gamma}{\partial y},
\label{eq:fp1} 
\end{equation} 
where 
\begin{equation}
\Gamma = \left\{ \begin{array}{cc}
\displaystyle a e^{-2y/3} P - \frac{\partial}{\partial y}\left(
be^{-2y/3}P\right) & \mbox{  if $y_d < y < 0$} \\
\displaystyle a e^{-2y_d/3} P - be^{-2y_d/3}\frac{\partial P}{\partial y}
 & \mbox{ if $y < y_d $} \\
\end{array}
\right.,
\label{eq:defgamma} 
\end{equation} 
is the probability flux, and 
\begin{equation}
y_d = \ln\left(\frac{d}{l_B}\right),
\label{eq:defyd} 
\end{equation} 
which is the value of $y$ at the dissipation scale.
We solve equation~(\ref{eq:fp1}) with initial condition $P(y) = \delta
(y-y_0)$ at $l=0$ and boundary conditions $P = 0 $ at $y=0$, $P \rightarrow 0$
as $y\rightarrow -\infty$, and $P$ and $\Gamma$ continuous at $y=y_d$.
For electron thermal conduction in galaxy clusters, the quantity of
interest is $L_{\rm S}$ when the initial separation $r_0$ is the electron
gyroradius, and thus we take $y_0 < y_d$.  The boundary condition
$P=0$ at $y=0$ means that $\Gamma(l, y=0)dl$ gives the probability
that a field-line pair separates to a distance~$l_B$ for the first
time after a distance between~$l$ and~$l+dl$ along the field.

We proceed by making the substitution
\begin{equation}
P = x^m f,
\label{eq:deff} 
\end{equation} 
with
\begin{equation}
x= e^{y/3}
\label{eq:defx}
\end{equation} 
and
\begin{equation}
m= 2 + 3a/2b.
\label{eq:defm} 
\end{equation}
We then take the Laplace transform of equation~(\ref{eq:fp1}), with
the Laplace transform of~$f$ defined by 
\begin{equation}
\overline{ f}(s) = \int_0^{\infty} f(l) e^{-sl} dl.
\label{eq:deflaplace} 
\end{equation} 
For $x_d < x < 1$, where 
$x_d = e^{y_d/3}$ is the value of $x$ at the dissipation scale,
\begin{equation}
\frac{\partial ^2 \overline{f}}{\partial x^2} + \frac{1}{x} \, \frac{\partial \overline{f}}{\partial x}
- \left(\frac{\nu^2}{x^2} + \frac{9s}{b}\right) \overline{f} = 0, 
\label{eq:fi} 
\end{equation} 
with
\begin{equation}
\nu = \frac{3a}{2b}.
\end{equation}
Since $f(1) = 0$, 
\begin{equation}
\overline{ f} = c_1[I_\nu(\psi x) K_\nu(\psi) - K_\nu(\psi x) I_\nu(\psi)],
\label{eq:fbessel} 
\end{equation} 
where $I_\nu$ and $K_\nu$ are modified Bessel's functions, 
\begin{equation}
\psi = 3\sqrt{\frac{s}{b}},
\label{eq:defalpha} 
\end{equation} 
and $c_1$ is a constant to be determined by applying the boundary
conditions at $x_d$ after the solution for $\overline{ f}$ for $x< x_d$ has been
obtained.

For $x< x_d$, 
\begin{equation}
\frac{\partial ^2 \overline{ f}}{\partial x^2} + \frac{5}{x} \,\frac{\partial \overline{ f}}
{\partial x} + \frac{(4-\nu^2 - \chi)\overline{ f}}{x^2} = -\frac{3x_d^2 x_0^{-m-1}\delta(x-x_0)}{b},
\label{eq:f2} 
\end{equation} 
where
\begin{equation}
\chi = \frac{9x_d^2 s}{b}
\label{eq:defchi} 
\end{equation} 
and 
\begin{equation}
x_0 = e^{y_0/3}.
\label{eq:defx0} 
\end{equation} 
For $x < x_0$,
\begin{equation}
\overline{ f} = c_2 x^{-2 + \sqrt{\nu^2
+ \chi}} + c_3 x^{-2 - \sqrt{\nu^2
+ \chi}} .
\label{eq:f3} 
\end{equation} 
For $x_0 < x < x_d$, 
\begin{equation}
\overline{ f} = c_4 x^{-2 + \sqrt{\nu^2 +
\chi}} + c_5 x^{-2 - \sqrt{\nu^2 + \chi}}.
\label{eq:f5}  
\end{equation} 
For  $Re(s) \geq 0$, the boundary
condition at~$y= -\infty$ implies that~$c_3 = 0$.  Integrating
equation~(\ref{eq:f2}) from~$x_0 - \epsilon$ to~$x_0 + \epsilon$ yields the 
jump condition for $\partial \overline{ f}/\partial x$ at~$x=x_0$.
After applying this jump condition and the continuity of~$f$ and~$\Gamma$ 
at $x=x_d$, we find that
\begin{equation}
c_1 = \frac{3 x_0^{-\nu +\sqrt{\nu^2 + \chi}} x_d^{-\sqrt{\nu^2 + \chi}}}
{b(hT - \psi x_d U)},
\label{eq:c1} 
\end{equation} 
where
\begin{equation}
T = I_\nu(\psi x_d) K_\nu(\psi) - K_\nu(\psi x_d) I_\nu(\psi),
\label{eq:defT} 
\end{equation} 
and
\begin{equation} 
U = I_\nu^\prime(\psi x_d)K_\nu(\psi) - K_\nu^\prime(\psi x_d)
I_\nu(\psi).
\label{eq:defu} 
\end{equation}

Since
$I_\nu^\prime (\psi)
K_\nu(\psi) - K_\nu^\prime(\psi)I_\nu(\psi) = 1/\psi$,
we find that
\begin{equation}
\overline{ \Gamma}(x=1) = - \frac{bc_1}{3}.
\label{eq:gamma1} 
\end{equation} 
Since
\begin{equation}
\langle l^n \rangle =  \left.\int_0^\infty \,dl\, l^n\, \Gamma\;\right|_{x=1}
= \left.\left(\frac{\partial }{\partial s}\right)^n 
\overline{ \Gamma}\;\right|_{x=1,s=0},
\end{equation} 
where $n$ is a non-negative integer, we have
\begin{equation}
\langle l^n \rangle = \left. - \frac{b}{3} \frac{\partial ^n c_1}{\partial s^n}\;\right|
_{s=0}.
\label{eq:lmom} 
\end{equation} 
Thus,
\begin{equation}
\langle l \rangle = \frac{9}{b} \left[
\frac{1}{4(\nu+ 1)} + \frac{x_d^2}{2\nu} \ln \left(\frac{x_d}{x_0}\right)
- \frac{x_d^2}{4\nu^2}\left(\nu - 1 + \frac{x_d^{2\nu}}{\nu + 1} \right) \right].
\label{eq:avl1} 
\end{equation} 
For fixed positive~$a$ in the limit~$x_d \rightarrow 0$, equation~(\ref{eq:avl1}) gives 
\begin{equation}
\langle l \rangle = \frac{9}{2(2b + 3a)}.
\label{eq:averagel} 
\end{equation} 
The full equation for $\langle l^2\rangle $ is very long and will not
be quoted here. However, for fixed positive~$a$
in the limit~$x_d\rightarrow 0$,  we find that 
\begin{equation}
\langle l^2 \rangle  = \frac{81 (3a + 6b)}{4(3a+2b)^2(3a+4b)},
\label{eq:averagel2} 
\end{equation} 
with  $L_{\rm S}/l_B = \langle l^2 \rangle /\langle l \rangle$ 
given by 
\begin{equation}
\frac{L_{\rm S}}{l_B} = \frac{9(3a + 6b)}{2(3a+2b)(3a+4b)}.
\label{eq:lrr} 
\end{equation} 
From equation~(\ref{eq:avl1}) 
$\langle l \rangle$ diverges for fixed~$x_d$
as~$a\rightarrow 0$  (i.e. $\nu \rightarrow 0$).

\begin{figure*}[h]
\vspace{8.5cm}
\includegraphics{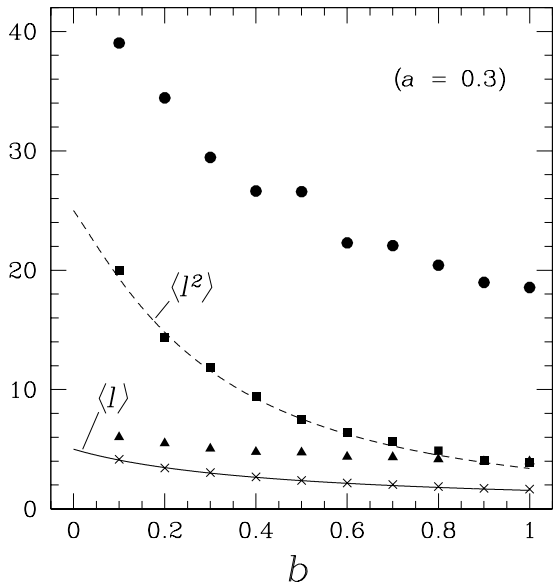}
\caption{The solid line gives $\langle l \rangle$ from
equation~(\ref{eq:averagel}). The dashed line gives $\langle l^2
\rangle$ from equation~(\ref{eq:averagel2}). The $\times$s and squares
give the values of $\langle l \rangle$ and $\langle l^2 \rangle$ from
Monte Carlo calculations with $\varphi = 10^{-4}$.  The triangles and
circles give the values of $\langle l \rangle$ and $\langle l^2
\rangle$ from Monte Carlo calculations with $\varphi = 1$.
For all data in plot, $a=0.3$.
\label{fig:comp0}}
\end{figure*}

We now check the analytic results with a Monte Carlo solution of
equation~(\ref{eq:fp0}) with initial conditions $P(y, l=0) = \delta
(y-y_0)$, and with $y_0 = -10$ and $y_d = -8$. The Monte Carlo
solution consists of iteratively incrementing a pair of numbers $(l,
y)$.  During each step, we increase~$l$ by an amount $\delta l =
e^{2y/3}\,\varphi$ (or $\delta l =e^{2y_d/3}\,\varphi$ if $y< y_d$) and
increase~$y$ by an amount $ \delta y = a \varphi \pm k$, where the
$\pm$~sign is determined randomly with equal chance for either
sign, and $\varphi$ is a constant. As $\varphi \rightarrow 0$,
each step becomes infinitesimal as is assumed in the Fokker-Planck
equation. The value of $k$ is chosen so that $\langle(\delta y)^2\rangle =
2b\varphi$ (i.e., $k = \sqrt{2b\varphi - a^2 \varphi ^2}$). We stop
incrementing $l$ and $y$ once $y$ reaches~$0$, and record the value
of~$l$ at the final step. We repeat this for 2000 ordered pairs to
obtain $\langle l \rangle$ and $\langle l^2 \rangle$. The results of
this procedure with $\varphi = 10^{-4}$ and $a= 0.3$ for various values
of~$b$ are shown in figure~\ref{fig:comp0}, along with the analytic
results of equations~(\ref{eq:averagel}) and (\ref{eq:averagel2}).  We
also plot Monte Carlo results with $\varphi =1$, which provide a
measure of the error associated with using a Fokker-Planck equation to
describe the discrete stochastic process described by
equations~(\ref{eq:defdell}), (\ref{eq:defa}), and~(\ref{eq:defb}).

\begin{figure*}[h]
\vspace{8.5cm}
\includegraphics{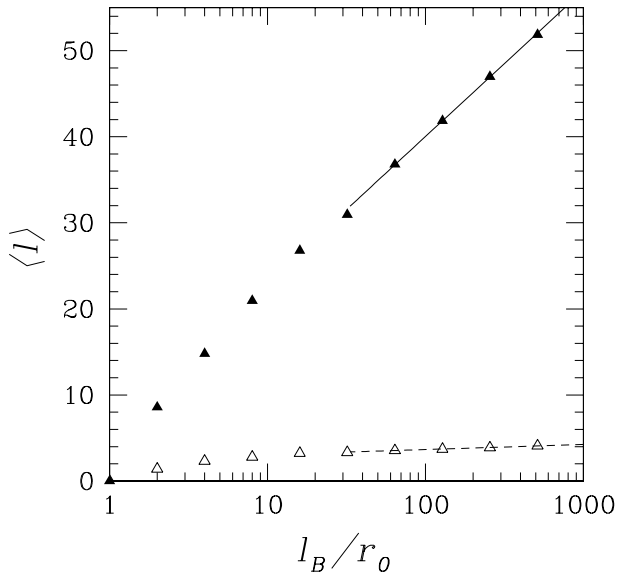}
\caption{Monte Carlo calculations
of $\langle l \rangle$ with~$b=0.17$, $l_B/l_d = 50$,
$\varphi=10^{-4}$ and two values of~$a$: $a=0.01$ (solid
triangles) and $a=0.29$ (open triangles). The solid
line is the analytic result [equation~(\ref{eq:avl1})]
with~$a=0.01$, and the dashed line is the analytic
result with~$a=0.29$.
\label{fig:f3}}
\end{figure*}

In figure~\ref{fig:f3} we plot Monte Carlo calculations of~$\langle l
\rangle$ as a function of initial field-line separation~$r_0$
for~$b=0.17$, $l_B/l_d = 50$, $\varphi=10^{-4}$, and two values
of~$a$: $a=0.01$ (solid triangles) and~$a=0.29$ (open triangles). The
solid line is a plot of equation~(\ref{eq:averagel}) for~$a=0.01$, and
the dashed line is a plot of equation~(\ref{eq:averagel})
for~$a=0.29$.  The figure shows the increase in~$\langle l \rangle$
as~$a\rightarrow 0$ for fixed~$x_d$.

\section{Direct numerical simulations}
\label{sec:comp} 

In this section, we study field-line separation in a static magnetic
field using data obtained from direct numerical simulations of MHD
turbulence. Galaxy clusters have very small mean magnetic fields,
little rotation and thus little helicity, and very large magnetic
Prandtl number~$P_m$, where~$P_m = \nu/\eta$, $\nu$ is the
viscosity,\footnote{For clusters with anisotropic plasma viscosity,
$\nu$ is taken to be the parallel viscosity.}  and~$\eta$ is the
resistivity. Ideally, we would like to generate cluster-like magnetic
fields self-consistently within a direct numerical simulation.
However, at present it is not clear how to do this.  Numerical
simulations of turbulent dynamos in high-$P_m$ plasmas driven by
non-helical forcing find that amplified magnetic fields remain
concentrated on very small (resistive) spatial scales (Maron \& Cowley
2001).  Yet the Faraday rotation produced by intracluster plasmas
indicates that clusters possess considerable amounts of magnetic
energy on large scales of order~$1-10$~kpc.  We are thus faced with
several alternatives. We can choose a numerical model that matches the
helicity, mean magnetic field, and magnetic Prandtl number conditions
of clusters, in which case the spatial scale of the model magnetic
field is far too small. Or we can choose a numerical model that
produces a large-scale field by, for example, including a mean
magnetic field, using helical forcing, or choosing initial conditions
that ensure that the magnetic field remains large-scale.\footnote{
Haugen et~al (2003) found that non-helical dynamos with $P_m$ up to~30
result in a magnetic power spectrum~$E_b(k)$ proportional
to~$k^{-1}$. Such a spectrum is neither large-scale-dominated nor
small-scale-dominated, since each logarithmic interval of order unity
in $k$-space contains the same amount of energy.  However, they
suggest that when both the ordinary Reynolds number~$R$ and~$P_m$ are
large, there is a $k^{-5/3}$ spectrum at large scales followed by
a~$k^{-1}$ spectrum at smaller scales, with the magnetic energy
dominated by large-scale fluctuations. If this suggestion is correct,
a very-high-resolution non-helical dynamo simulation with large~$R$
and large~$P_m$ would be a self-consistent way to generate a
large-scale magnetic field in cluster-like conditions. Such a
simulation, however, would be beyond our current computational
resources.}

In this paper, we follow the latter course. We initialize a simulation
with a random-phase magnetic field containing some amount of magnetic
helicity, as well as random velocities with kinetic energies
comparable to the magnetic energy. We allow the system to decay,
leaving behind a large-scale magnetic field, which contains 10\% of
the maximum magnetic helicity for that magnetic energy. We then force
the system non-helically at wavenumbers between~$2\pi/L_{\rm box}$
and~$4\pi/L_{\rm box}$, where~$L_{\rm box}^3$ is the volume of the
simulation cube. The forcing sustains a Kolmogorov-like spectrum of
magnetic and kinetic energy. The result is a turbulent magnetic field
that is dominated by large-scale fluctuations, has a Kolmogorov-like
inertial range extending to small scales, has zero mean, and has
relatively little magnetic helicity, which we take to be a reasonable
model for magnetic fields in clusters. We carry out a first simulation
(simulation~A1) on~$256^3$ grid points, and then use A1 as an initial
condiction for a higher resolution version (simulation~A2) on~$512^3$
grid points with reduced resistivity and viscosity. We use
incompressible simulations, which are a reasonable approximation for
subsonic turbulence in clusters. The simulations are three-dimensional
and periodic, and the pseudo-spectral numerical method is described by
Maron \& Goldreich (2001).  We use Newtonian viscosity and resistivity
with $P_m=1$, since simulating $P_m \gg 1$ in our isotropic-viscosity
simulations requires a large viscosity that damps small-scale Alfv\'en
waves; small-scale Alfv\'enic turbulence is present in clusters due to
anisotropic plasma viscosity (Goldreich \& Sridhar 1995, Quataert
1998), and plays an important role in field-line separation.  In a
companion paper, Maron, Chandran, \& Blackman (2003), we study
field-line separation in numerical simulations of different types of
MHD turbulence and dynamo-generated fields.

\begin{figure}[h]
\vspace{9cm}
\includegraphics{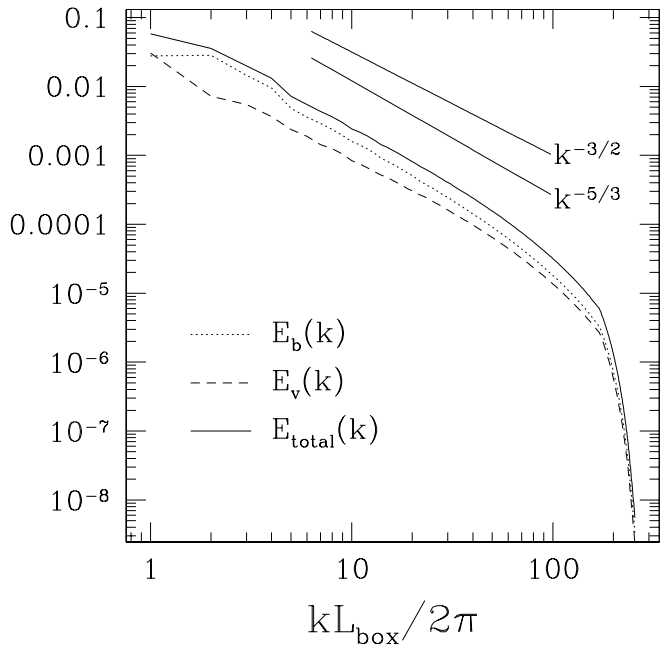}
\caption{Power spectra in simulation~A2. 
The dotted line is the magnetic power spectrum~$E_b(k)$, the dashed line is 
the kinetic power spectrum~$E_v(k)$, and the solid
line is the total-energy spectrum~$E_{\rm total} = E_v(k) + E_b(k)$.
\label{fig:f4}}
\end{figure}

The simulation parameters for~A1 and~A2 are summarized in
table~\ref{tab:t1}.  In figure~\ref{fig:f4} we plot time averages of
the magnetic power spectrum~$E_b(k)$ [the total magnetic energy is
$\int E_b(k) dk$], the kinetic power spectrum~$E_v(k)$ [the total
kinetic energy is $\int E_v(k) dk$], and the total-energy
spectrum~$E_{\rm total}(k) = E_v(k)+E_b(k)$ in simulation~A2.  We set
$\pi/l_B$ equal to the maximum of $k E_b(k)$, giving~$l_B=0.25L_{\rm box}$,
and we set $\pi/l_d$
equal to the maximum of $k^3E_b(k)$.

\begin{table}[h]
\begin{tabular}{ccccccc}
\hline
\hline 
&&&&&&
\vspace{-0.35cm} 
\\
Simulation & Grid points & $\displaystyle
|\langle{\bf B}\rangle| $ & $H_m$ & $l_B/l_d$ & $\alpha$ 
 \vspace{0.05cm} & $P_m = \nu/\eta$ \\
\hline  
\vspace{-0.4cm} \\
A1 & $256^3$ & 0  & 0.1 & \hspace{0.2cm}  23 \hspace{0.2cm} & 2.4  & 1\\
A2 & $512^3$ & 0 & 0.1 & 50 & 2.4 & 1 \\
\hline
\hline
\end{tabular}

\caption{$|\langle {\bf B} \rangle|$ is the strength of
the mean magnetic field, $H_m$ is the magnetic helicity
divided by the maximum possible magnetic helicity at that level of
magnetic energy, $l_B/l_d$ is the ratio of outer scale to inner scale,
$\alpha$ is the single-field-line diffusion coefficient in
equation~(\ref{eq:tr}), and $P_m=\nu/\eta$ is the magnetic Prandtl
number, where~$\nu$ and~$\eta$ are the viscosity and resistivity.
\label{tab:t2} }
\end{table}

We run each simulation until the power spectrum reaches a statistical
steady state before we start analyzing field lines.
For simulation~A1, we take seventeen snapshots of
the magnetic field separated in time by an interval~$0.4l_B/u$,
where~$u$ is the rms turbulent velocity.  To calculate~$L_{\rm S}$ and
$\langle l \rangle$ for an initial field-line separation~$r_0$, we
introduce into each snapshot of the magnetic field 2,000 pairs of
field-line tracers whose initial separation vector~${\bf r}_0$ is
perpendicular to the local field. We use linear interpolation to
obtain the magnetic field between grid points, employ second-order
Runge-Kutta to integrate field-lines, and iteratively reduce the
length step in the field-line integrations to achieve convergence. For
simulation~A2, we carry out the same procedure, but we use five
snapshots of the magnetic field separated in time by an interval~$0.2
l_B/u$, and we use 20,000 pairs of~field-line tracers per snapshot.

\begin{figure}[h]
\vspace{9cm}
\includegraphics{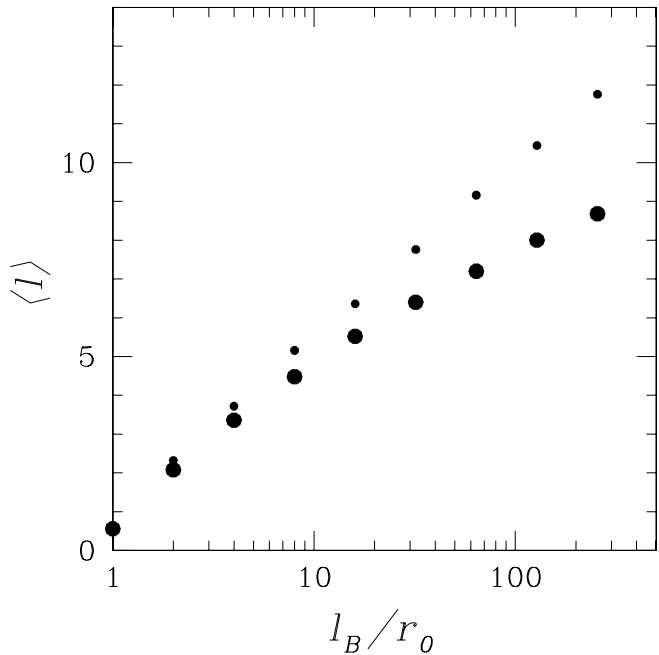}
\caption{The average distance in units of~$l_B$
that a field-line pair must be followed before separating by a
distance~$l_B$, denoted~$\langle l \rangle$, as a function of
initial field-line separation~$r_0$ for simulations~A1 (small
circles) and~A2 (large circles).
\label{fig:f5}}
\end{figure}

\begin{figure}[h]
\vspace{9cm}
\includegraphics{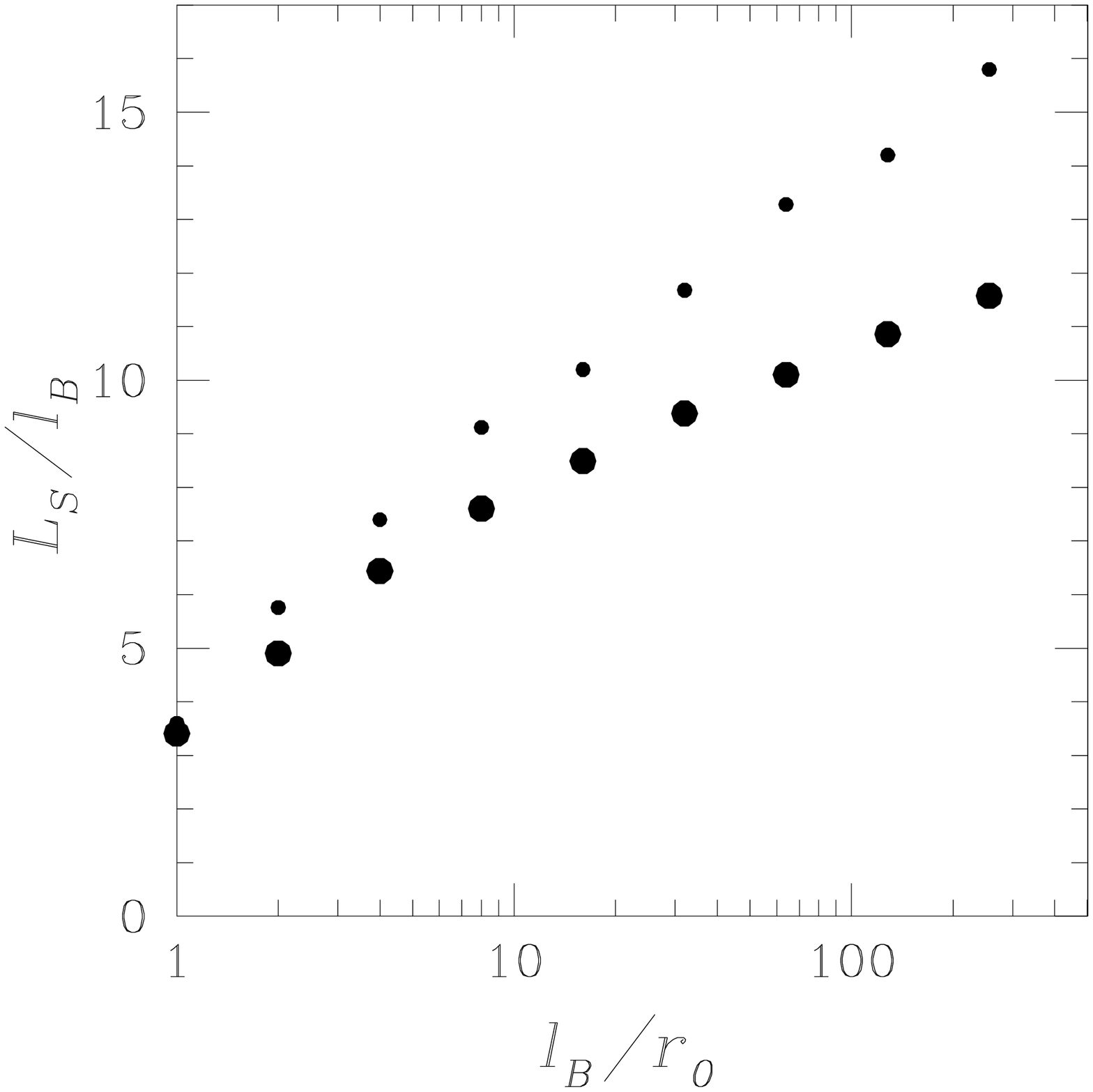}
\caption{Value of~$L_{\rm S}$ as a function of initial field-line
separation~$r_0$ in simulations~A1 (small circles) and~A2 (large
circles).
\label{fig:f6}}
\end{figure}

We first seek to test the qualitative prediction of the Fokker-Planck
model and previous theoretical treatments (Jokipii 1973, Skilling
et~al 1974, Narayan \& Medvedev 2001) that~$\langle l \rangle$
and~$L_{\rm S}/l_B$ asymptote to a constant of order a few as~$r_0$ is
decreased towards~$l_d$ in the large-$l_B/l_d$ limit. In figure~\ref{fig:f5}, we
plot~$\langle l \rangle$ for simulation~A1  and
simulation~A2. The lower-resolution data of
simulation~A1 suggest the scaling~$\langle l \rangle \propto \ln
(l_B/r_0)$ for $l_d < r_0 <0.25 l_B$, in contradiction to the
theoretical treatments. On the other hand, for simulation~A2,
the curve through the data is concave downward for~$l_d < r_0 <
l_B$. Moreover, figure~\ref{fig:f5} shows that the
simulation~A1 data, and probably also the A2~data, have not converged
to the high-Reynolds-number values of~$\langle l \rangle$ 
for~$l_B/16 < r_0 < l_B$, values of~$r_0$ that are within the
inertial ranges ($l_d$ to $l_B$) of both simulations. Also,
the slope $d\langle l \rangle/d(\ln(l_B/r_0))$ for both~$r_0<l_d$
and~$r_0<l_B/10$ 
decreases significantly when $l_B/l_d$ is doubled. A comparison
of the data for~A1 and~A2 thus suggests that in the large-$l_B/l_d$ limit
$\langle l \rangle$ asymptotes to a value of order several~$l_B$
as~$r_0$ is decreased towards~$l_d$, as in the Fokker-Planck
model and previous studies. The same comments apply to
the data for~$L_{\rm S}$, which are plotted in figure~\ref{fig:f6}.

\begin{figure*}[h]
\vspace{9cm}
\includegraphics{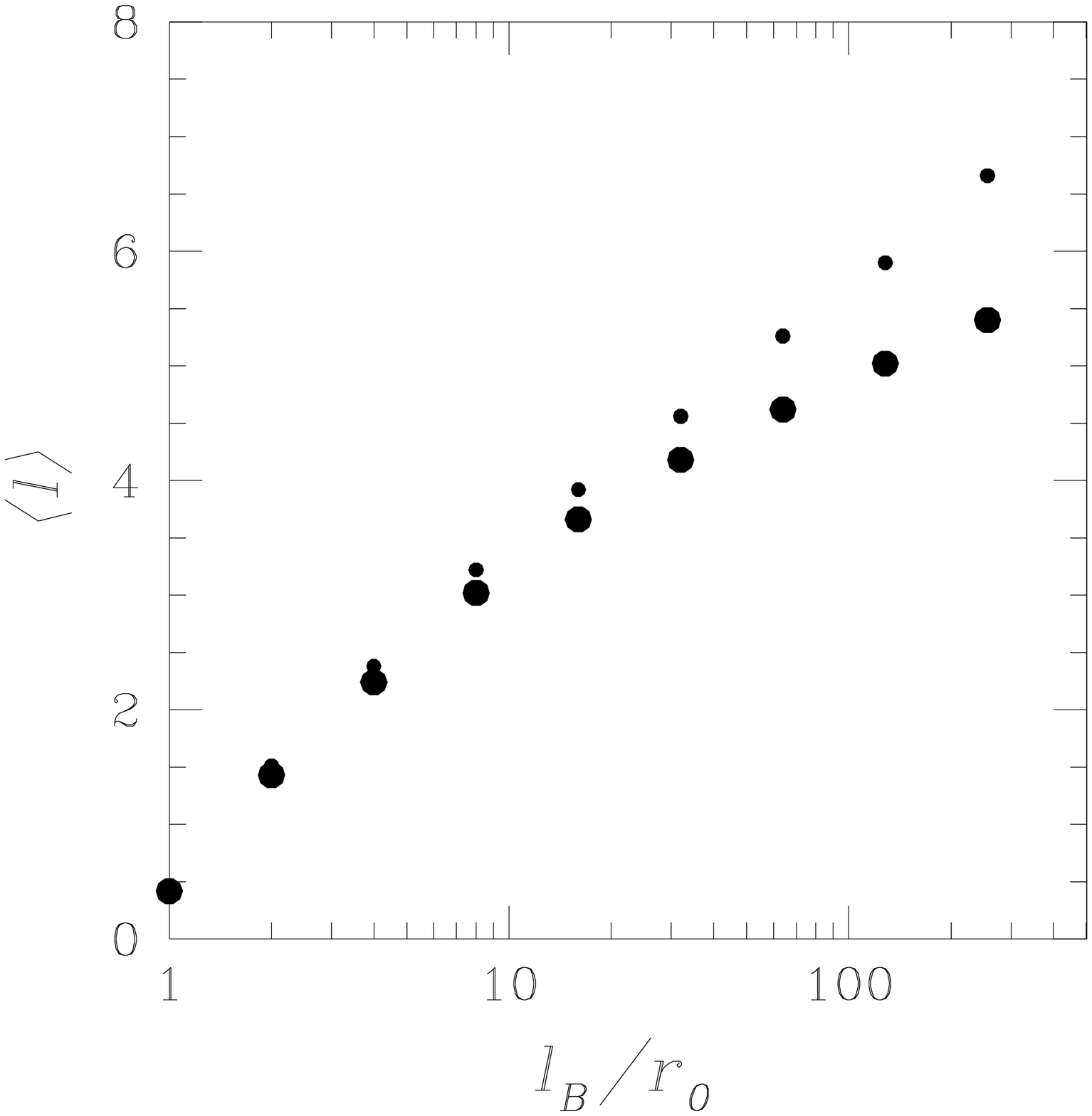}
\caption{The average distance in units of~$l_B$
that a field-line pair must be followed before separating by a
distance~$l_B$ as a function of
initial field-line separation~$r_0$ for simulations~A1 and~A2 (small
and large solid circles). Same as figure~\ref{fig:f5}, except
that~$l_B$ is redefined to be twice as large.
\label{fig:f7}}
\end{figure*}

\begin{figure*}[h]
\vspace{9cm}
\includegraphics{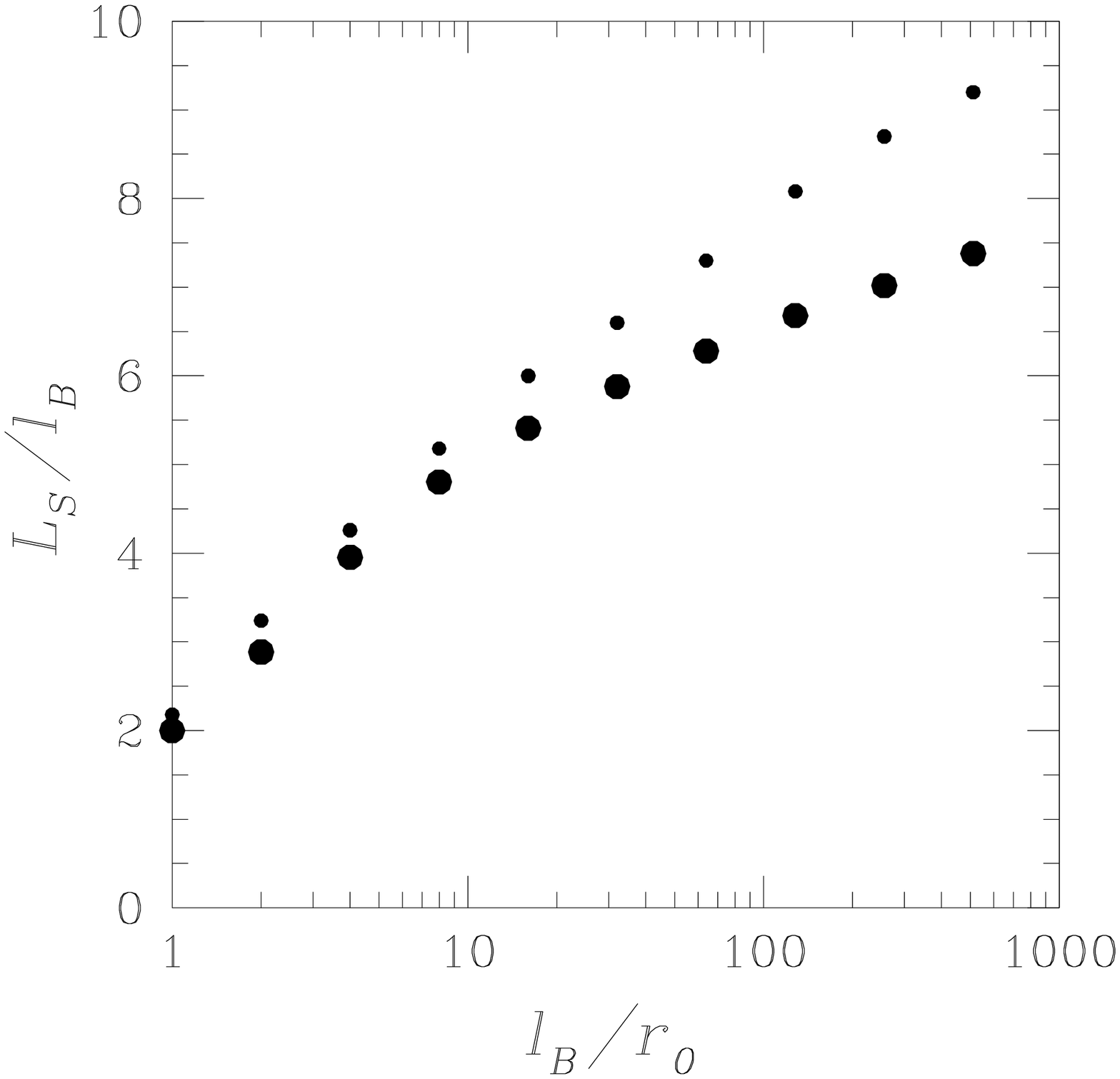}
\caption{Value of~$L_{\rm S}$ as a function of initial field-line
separation~$r_0$ in simulations~A1 (small circles) and~A2 (large
circles). Same as figure~\ref{fig:f6}, except that~$l_B$
is redefined to be twice as large. 
\label{fig:f8}}
\end{figure*}

Because the definition of~$l_B$ is not unique, we recalculate
$\langle l \rangle$ and $L_{\rm S}$, setting $2\pi/l_B$
equal to the maximum of~$kE_b(k)$, so that~$l_B$ is twice
its former value. Note that this affects both the unit for
measuring distance along the field and also the distance
to which field lines must separate. We plot the
results in figures~\ref{fig:f7} and~\ref{fig:f8}, which
are qualitatively similar to the results based on our
original definition of~$l_B$.

We note that for $r_0 = l_B$, $\langle l \rangle$ and~$L_{\rm S}$ are
by definition~0.  The numerical-simulation data points that appear to
be plotted above~$l_B/r_0 = 1$ actually correspond to~$r_0$ just slightly
smaller than~$l_B$, indicating that $\langle l \rangle$
and~$L_{\rm S}$ are discontinuous at $r_0 = l_B$ in the numerical
simulations. The reason is that for $r_0 $ just slightly less than
$l_B$, some fraction of the field line pairs are initially converging
and must be followed a significant distance before they start to
diverge.

\begin{figure}[h]
\vspace{9cm}
\includegraphics{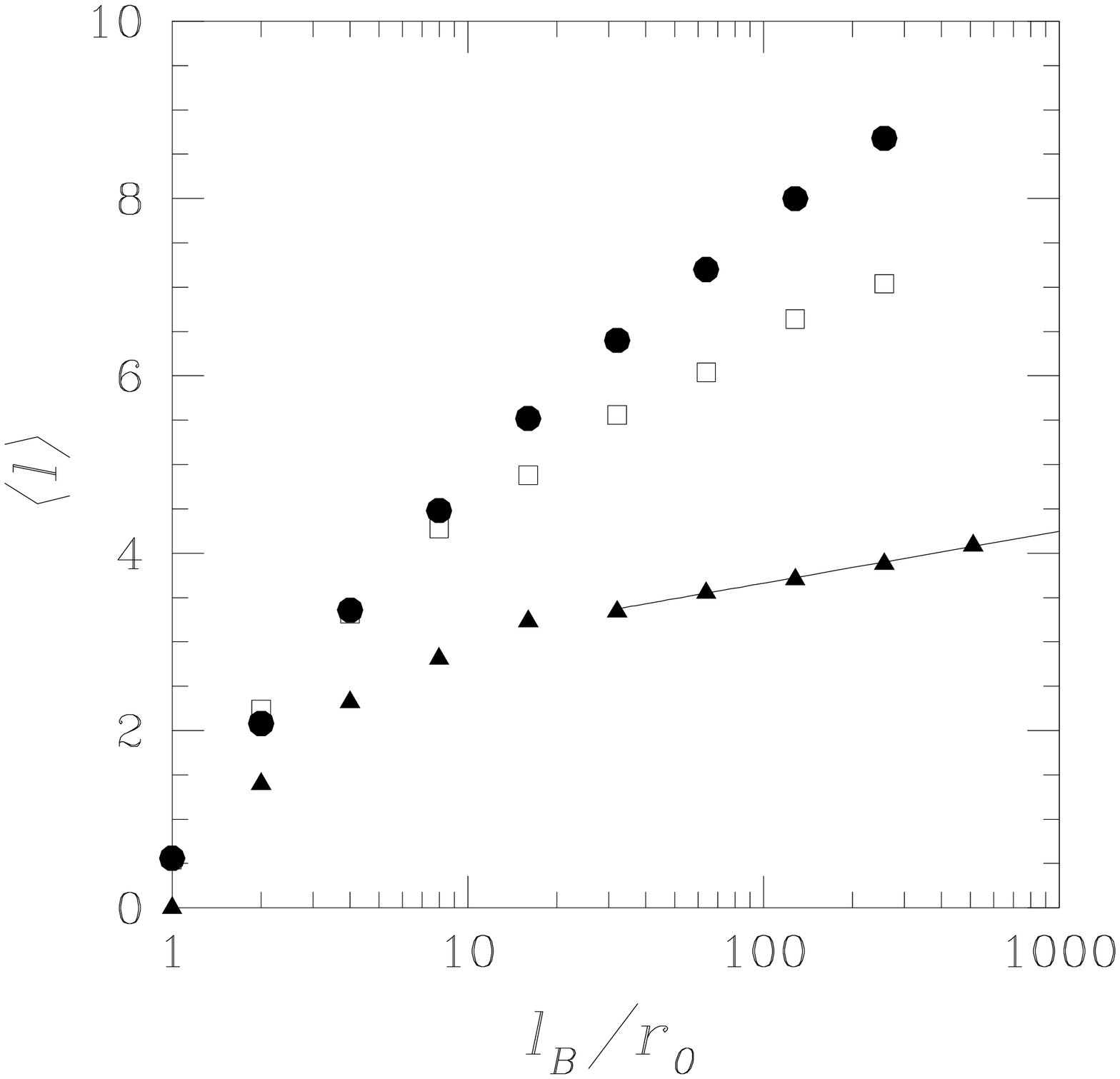}
\caption{The average distance in units of~$l_B$
that a field-line pair must be followed before separating by a
distance~$l_B$ as a function of
initial field-line separation~$r_0$ for simulation~A2 (solid circles),
the random-phase version of~A2 (open squares),
Monte Carlo simulations with~$\varphi = 10^{-4}$
(solid triangles),
and the analytic Fokker-Planck model (solid line).  The Monte Carlo
and Fokker-Planck solutions use $a=0.29$, $b=0.17$, and
$l_B/l_d =50$, values corresponding to simulation~A2.
\label{fig:f9}}
\end{figure}

We evaluate characteristic values of $a$ and~$b$, defined in
equations~(\ref{eq:defa}) and (\ref{eq:defb}), in simulation~A2 by
calculating the mean and mean-square increments to~$y= \ln(r/l_B)$ for
field-line pairs initially separated by a distance $l_B/8$ during a
displacement of~$l_B/4$ along the magnetic field, using our original
definition of~$l_B$ [$\pi /l_B$ = maximum of~$kE_b(k)$]. We find
that~$a=0.29$ and~$b=0.17$.  These values are used to obtain the
Fokker-Planck results plotted in figure~\ref{fig:f9}.  The Monte Carlo
results in figure~\ref{fig:f9} use the same values of~$a$ and~$b$ and
employ many small random steps~($\varphi=10^{-4}$), and are thus
expected to reproduce the Fokker-Planck results.  We also
plot~$\langle l \rangle$ for a random-phase version of~A2, which is
obtained by assigning each Fourier mode in simulation~A2 a random
phase without changing the modes' amplitudes. The figure shows
that~$\langle l \rangle$ is moderately larger in the direct numerical
simulations than in both the random-phase data and the Fokker-Planck
model.  If we take $a=0.29$, $b=0.17$, $l_d = \rho_i = 43\rho_e$, and
$l_B/l_d \gg 1$, the Fokker-Planck model yields $L_{\rm S}(\rho_e)
\simeq L_{\rm S}(l_d) \simeq 4.5 l_B$ and the Monte Carlo model with
order-unity random increments to~$\ln(r/l_B)$ (i.e.~$\varphi = 1$)
yields $L_{\rm S}(\rho_e) \simeq L_{\rm S}(l_d) \simeq 6.5 l_B$.

\begin{figure*}[h]
\vspace{9cm}
\includegraphics{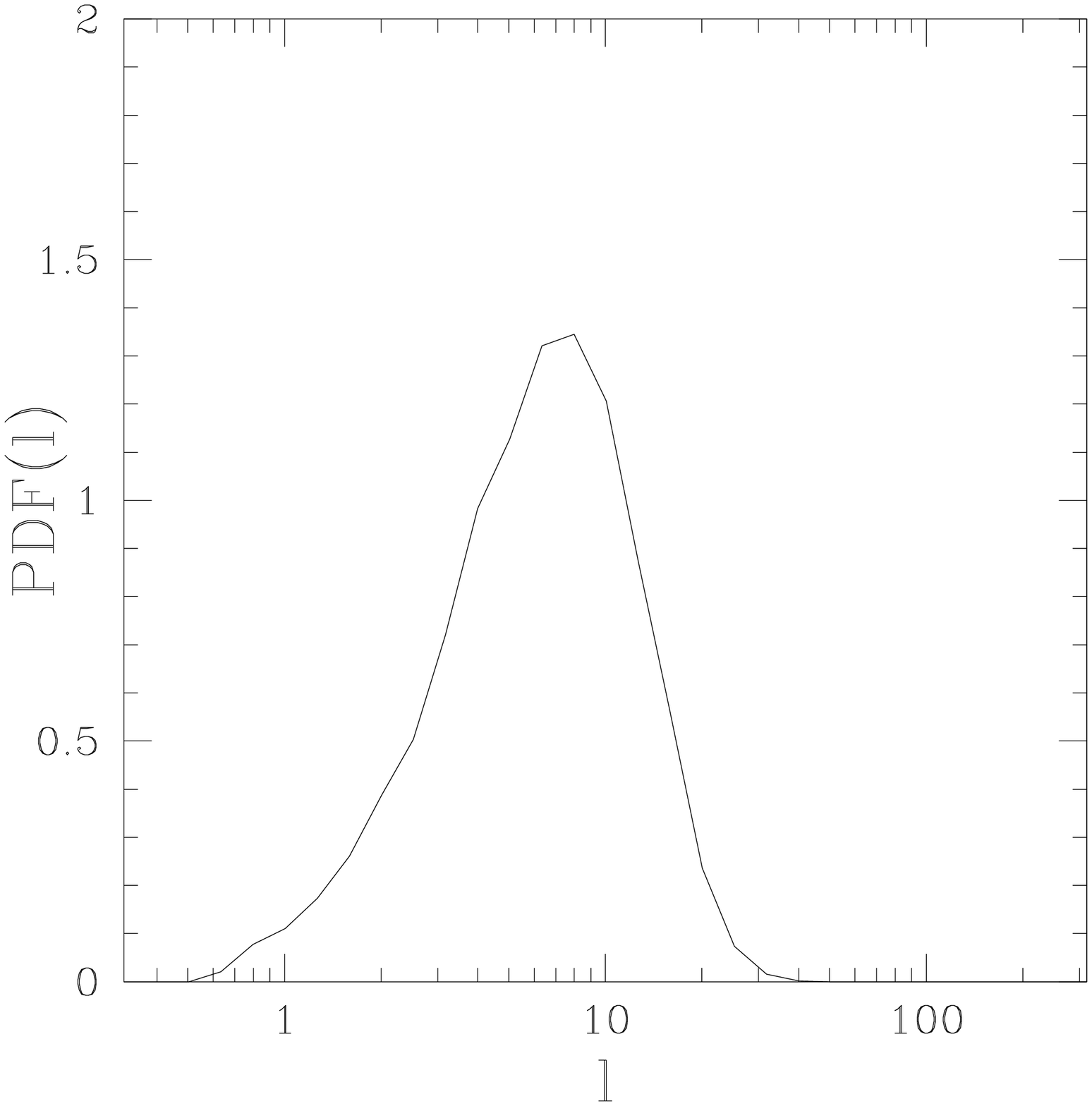}
\caption{Probability distribution of the distance (in units of~$l_B$)
a field-line pair in simulation~A2 initially separated by a
distance~$l_d$ must be followed before separating by a distance~$l_B$.
\label{fig:f10}}
\end{figure*}

In figure~\ref{fig:f10} we plot the probability distribution of the
distance in units of~$l_B$ that a pair of field lines in simulation~A2
with initial separation~$r_0 = l_d$ must be followed before the field
lines separate by a distance~$l_B$.  We define the function~PDF($l$)
so that the probability that~$l$ lies in some interval is proportional
to the corresponding area under the plotted curve.

For clusters, $l_B/l_d \simeq l_B/\rho_i \simeq 10^{13}$.  In the
large-$l_B/l_d$ limit, the numerical simulations and theoretical
models indicate that $L_{\rm S}$ asymptotes to a value of order
several~$l_B$ as~$r_0$ is decreased towards~$l_d$, and $L_{\rm S}$ is
not expected to increase appreciably as $r_0$ is further decreased
from~$l_d=\rho_i$ to~$\rho_e$. Thus, $L_{\rm S}(\rho_e) \simeq L_{\rm
S}(l_d)$.  To estimate $L_{\rm S}(l_d)$ in clusters, we note
that~$L_{\rm S}(l_d) \simeq 11 l_B$ in simulation~A1, and $L_{\rm
S}(l_d) \simeq 10 l_B$ in simulation~A2.  When the definition of~$l_B$
is changed as in figure~\ref{fig:f8}, so that $2\pi/l_B$ corresponds
to the maximum of~$k E_b(k)$, then $L_{\rm S}(l_d) \simeq 7 l_B$ in
simulation~A1 and $L_{\rm S}(l_d) \simeq 6.5 l_B$ in simulation~A2.
As mentioned previously, it is not clear which definition of~$l_B$
leads to a more accurate prediction of~$\kappa_T$. We conclude
from the direct numerical simulations suggest that $L_{\rm S}(\rho_e)
\simeq L_{\rm S}(l_d) \simeq 5-10 l_B$ in the large-$l_B/l_d$ limit.

\section{Thermal conduction in time-varying
turbulent magnetic fields}
\label{sec:rec} 

In this section, we develop a phenomenology of particle
diffusion in time-varying turbulent magnetic fields under the
questionable assumption that the magnetic field is completely
randomized and reconnected on the eddy turnover time~$\tau$ at
scale~$l_B$, $\tau = l_B/u$, where~$u$ is the rms velocity and the
velocity outer scale~$l_0$ is assumed equal to~$l_B$.  A similar
assumption was explored by Gruzinov (2002). For simplicity, we assume
that the magnetic-field randomization occurs instantaneously at
regular time intervals of duration~$\tau$. We assume $\tau \gg
\lambda/v_{\rm te}$, where~$v_{\rm te}$ is the electron thermal
velocity and $\lambda$ is the Coulomb mean free path, so that particle
motion along the field over a time interval~$\tau$ is diffusive.  We
assume that $\lambda \ll L_{\rm S}$, so that particle motion along the
field over a distance~$L_{\rm S}$ is diffusive.  We also assume
that~$v_{\rm te} \gg u$.

There are three limiting cases.  First, if $\tau \gg L_{\rm
S}^2/D_\parallel$, a particle escapes its initial field line through
parallel motion and slow cross-field diffusion before the field is
randomized.  The ``fundamental random-walk step'' is of length~$L_{\rm
S}$ along the field, as in section~\ref{sec:static}, and takes a
time~$L_{\rm S}^2/D_\parallel$. There are $n_{\rm steps} = \tau
D_\parallel /L_{\rm S}^2$ such steps during each time
interval~$\tau$. The three dimensional distance travelled by an
electron along the magnetic field during one ``fundamental random-walk
step'' is~$\sim \sqrt{L_{\rm S}l_B}$, and the three-dimensional
distance travelled during~$n_{\rm steps}$ steps is $\sim \sqrt{n_{\rm
steps} L_{\rm S} l_B}$.  On the other hand, a fluid element travels a
distance~$l_B$ during a time~$\tau$.  Since $L_{\rm S} > l_B$
and~$n_{\rm steps} \gg 1$, the distance travelled by a single electron
during a time~$\tau$ is much greater than the distance travelled by a
fluid element, and the fluid motion can be ignored.  The electron diffusion
coefficient is then the same as in the static-field case.

At the other extreme is the limit~$\tau \ll l_B^2/D_\parallel$. Since $l_B <
L_{\rm S}$, a particle escapes a field line through field-line
randomization before it escapes through parallel motion and slow
cross-field diffusion, and the fundamental random-walk step is of
duration~$\tau$. The distance an electron travels along the field due
to parallel diffusion, $\sqrt{D_\parallel \tau}$, is less than the
distance~$l_B$ that a fluid element travels.  The net displacement of
an electron (or ion) during one fundamental random step,
denoted~$\triangle r$, is then given by
\begin{equation}
\triangle r \sim l_B.
\end{equation} 
Since the field is completely randomized after a time~$\tau$, 
magnetic tension does not inhibit the wandering
of fluid parcels over times~$\gg \tau$, as in
Vainshtein \& Rosner (1991) and Cattaneo (1994).
Successive random steps are thus uncorrelated,
giving a diffusion coefficient $\triangle r^2/\tau \sim l_B^2/\tau
= u l_B$, as in hydrodynamical turbulent diffusion.
In this limit, the parallel diffusion of electrons plays
no role.

The third and intermediate case is 
$l_B^2/D_\parallel \ll \tau \ll L_{\rm S}^2/D_\parallel$.
In this case, the field is again randomized before a particle
can escape its initial field line through parallel motion
and slow cross-field diffusion,
and the fundamental random-walk step is of duration~$\tau$.
During one such step, a particle moves an rms distance~$\sim
\sqrt{D_\parallel \tau}$ along the field, which
corresponds to an rms three
dimensional displacement
\begin{equation}
\triangle r  \sim ( l_B \sqrt{D_\parallel
\tau})^{1/2}.
\end{equation} 
The diffusion coefficient $\triangle r^2/\tau$ is then
\begin{equation}
D \sim \sqrt{D_\parallel u l_B}.
\label{eq:d4} 
\end{equation}

As an example, we consider the cluster~A1795 at a distance of 100~kpc
from cluster center, with an electron density of~$0.01 \mbox{
cm}^{-3}$ and a temperature of~$5$~keV (Ettori et~al 2002). We
consider slightly superthermal electrons with~$D_\parallel =
\kappa_{\rm S} = 10^{31} \mbox{ cm}^2/\mbox{s}$.  Churazov
et~al (2003) find evidence for turbulent velocities of order one-half
the sound speed in the hot intracluster plasma of the Perseus cluster,
which we take to be typical of A1795 as well, giving~$u\simeq
350$~km/s.  We also assume~$l_B = 10$~kpc at this distance from
cluster center, and take~$L_{\rm S} = 6 l_B$.  For these parameters,
the plasma satisfies $l_B^2/D_\parallel < \tau < L_{\rm
S}^2/D_\parallel$.
Equations~(\ref{eq:defkt}), (\ref{eq:dpard0}),
and~(\ref{eq:d4}) then give
\begin{equation} 
\kappa_T\sim \sqrt{\kappa_{\rm S} u l_B},
\label{eq:kttr2} 
\end{equation} 
which is about $0.3\kappa_{\rm S}$, or $3 u l_B$. 
This value is roughly twice
the estimate~$\kappa_{\rm S} l_B/L_{\rm S}$, but given the
uncertainties in both estimates, and the untested assumption that
field lines are completely randomized and reconnected on the time
scale~$\tau$, it is not clear that turbulent resistivity in fact
enhances the thermal conductivity in clusters.

\section{Summary}
\label{sec:conc} 

In this paper we consider the effects of field-line tangling and
turbulent resistivity on the thermal conductivity~$\kappa_T$ in galaxy
clusters. In the static-magnetic-field approximation, tangled field
lines force electrons to move greater distances in traveling from
hotter regions to colder regions, reducing $\kappa_T$ by a factor
of~$\sim 5-10$ relative to the Spitzer thermal
conductivity~$\kappa_{\rm S}$ of a non-magnetized plasma for typical
cluster parameters.  It is possible that turbulent resistivity 
enhances~$\kappa_T$ by a moderate amount relative to
the static-field estimate for typical cluster conditions, but further
work is needed to investigate this possibility.

\acknowledgements
We thank Eric Blackman, Steve Cowley, and Alex Schekochihin for valuable discussions.
This work was supported by NSF grant AST-0098086 and DOE grants
DE-FG02-01ER54658 and DE-FC02-01ER54651 at the University of Iowa,
with computing resources provided by the National Partnership for Advanced
Computational Infrastructure and the National Center for Supercomputing
Applications.

\vspace{0.2cm} 

Allen, S., Taylor, G., Nulsen, P., Johnstone, R., David, L., Ettori, S., Fabian, A.,
Forman, W., Jones, C., \& McNamara, B. 2001, MNRAS, 324, 842

Binney, J., \& Cowie, L. 1981, ApJ, 247, 464

Binney, J., \& Tabor, G. 1995, MNRAS, 276, 663

Bohringer, H., \& Morfill, G. E. 1988, ApJ, 330, 609

Bregman, J., \& David, L. 1989, ApJ, 341, 49

Cattaneo, F. 1994, ApJ, 434, 200

Chandran, B. 1997, ApJ, 485, 148

Chandran, B. 2003, ApJ, submitted

Chandran, B., \& Cowley, S. 1998, Phys. Rev. Lett., 80, 3077

Chandran, B., \& Rodriguez, O. 1997, ApJ, 490, 156

Chandrasekhar, S. 1943, Rev. Mod. Phys., 15, 1

Cho, J., \& Lazarian, A., 2003 Phys. Rev. Lett. in press

Cho, J., Lazarian, A., Honein, A., Knaepen, B., Kassinos, S., and Moin P. 2003,
ApJ, submitted

Cho, J., \& Vishniac, E. 2000, ApJ, 539, 273 

Chun, E., \& Rosner, R. 1993, ApJ, 408

Churazov, E., Forman, W., Jones, C., Sunyaev, R., \& Bohringer, H. 2003,
MNRAS, accepted for publication

Ciotti, L., \& Ostriker, J. 2001, ApJ, 551, 131

Churazov, E., Sunyaev, R., Forman, W., \& Bohringer, H. 2002, MNRAS, 332, 729

Crawford, C., Allen, S., Ebeling, H., Edge, A., Fabian, A. 1999, MNRAS, 306, 875

Ettori, S., Fabian, A., Allen, S., \& Johnstone, R. 2002, MNRAS, 331, 635

Fabian, A. C. 1994, Ann. Rev. Astr. Astrophys., 32, 277

Fabian, A. C. 2002, astro-ph/0210150

Goldreich, P. \& Sridhar, S. 1995, ApJ, 438, 763

Gruzinov, A. 2002, astro-ph/0203031

Krommes, J., Oberman, C., \& Kleva, R. 1983, J. Plasma Phys., 30, 11

Jokipii, J. 1973, ApJ, 183, 1029

Lithwick, Y., \& Goldreich, P. 2001, ApJ, 562, 279

Loewenstein, M., \& Fabian, A. 1990, MNRAS, 242 120

Malyshkin, L., \& Kulsrud, R. 2001, ApJ, 549, 402

Maron, J., Chandran, B., \& Blackman, E. 2003, in preparation

Maron, J., \& Cowley, S. 2001, astro-ph/0111008

Maron, J., \& Goldreich, P. 2001, ApJ,  554, 1175

Maron, J., \& Blackman, E. 2002, ApJ, 566, L41

Narayan, R., \& Medvedev, M. 2001, ApJ, 562, 129

Pedlar, A., Ghataure, H. S., Davies, R. C., Harrison, B. A., Perley, R.,
Crane, P. C., \& Unger, S. W. 1990, MNRAS, 246, 477

Perley, R., Taylor, G. 1991, AJ, 101, 1623

Peterson, J. A. et~al 2001, A\&A, 365, L104

Pistinner, S., \& Eichler, D. 1998, MNRAS, 301, 49

Qin, G., Matthaeus, W., \& Bieber, J. 2002a, ApJ, 578, L117

Qin, G., Matthaeus, W., \& Bieber, J. 2002b, Geophys. Res. Lett. 29, 7

Quataert, E. 1998, ApJ, 500, 978

Rechester, R., \& Rosenbluth, M. 1978, Phys. Rev. Lett., 40, 38

Rosner, R., \& Tucker, W. 1989, ApJ, 338, 761

Skilling, J., McIvor, I., \& Holmes, J. 1974, MNRAS, 167, 87P

Tabor, G., \& Binney, J. 1993, MNRAS, 263, 323

Tamura, T. et~al 2001, A\&A, 365, L87

Tao, L. 1995, MNRAS, 275, 965

Taylor, G., Fabian, A., Allen, S. 2002, MNRAS, 334, 769  

Taylor, G., Govoni, F., Allen, S., Fabian, A. 2001, MNRAS, 326, 2

Tribble, P. 1989, MNRAS, 238, 1247

Tucker, W. H., \& Rosner, R. 1983, ApJ, 267, 547

Vainshtein, S., \& Rosner, R. 1991, ApJ, 376, 199

Voigt, L. M., Schmidt, R. W., Fabian, A. C., Allen, S. W., \& Johnstone, R. M. 2002,
{\em Mon. Not. R. Astr. Soc.}, submitted (astro-ph/0203312)

Zakamska, N., \& Narayan, R. 2003, ApJ, in press

\end{document}